\documentclass[prd,tightenlines,amsfonts,amssymb,amsmath]{revtex4}
\usepackage{amssymb}
\begin{document}

\title{Testing Chameleons Using Neutron Interferometry}

\author{Robert Poltis}
\email{robert.poltis@uct.ac.za}
\affiliation{Astrophysics, Cosmology and Gravitation Centre, Department of Mathematics and Applied Mathematics, University of Cape Town, Rondebosch 7701, South Africa}

\begin{abstract}
Chameleons are a well motivated scalar field that might explain the observed late time accelerated expansion of the universe. Chameleons possess the interesting property that their mass, and hence interaction range, is dependent on the density of their environment. One very appealing feature of chameleons is the potential to test a model of dark energy in a laboratory setting. Here we briefly review two proposed experiments to search for chameleons using neutron interferometry techniques for the Snowmass 2013 Community Summer Study.
\end{abstract}

\maketitle

In addition to constraining the chameleon-photon coupling through "light shining through a wall" and "trapping chameleons in a jar" in the GammeV and CHASE laboratory experiments \cite{Ahlers:2007st}-\cite{Upadhye:2012ar}, the chameleon-matter coupling may be also be investigated by means of neutron interferometry experiments. The neutron interferometer discussed in \cite{Pokotilovski:2012js} consists of two beams of cold neutrons traveling in a plane parallel to the ground (which we will denote the $x-z$ plane). The beams emanate from a slit located at ($x=0,z=a$), and recombine at ($x=L,z=b$). One beam travels from ($x=0,z=a$) to ($x=L,z=b$) directly, while the other beam bounces off a reflecting plane situated along $z=0$ before recombining with the first beam at ($x=L,z=b$) (see Fig.~(1) of \cite{Pokotilovski:2012js}).

The phase shift measured when the neutron beams combine contains contributions from several factors which we discuss here.
\begin{equation}
\psi=\psi_{geom}+\psi_{gr}+\psi_{refl}+\psi_{cor}+\psi_{cham}
\end{equation}
For an interferometer of interference coordinate b, The geometric phase shift between the two neutron beams with wave vector $k$ absent any potential is 
\begin{equation}
\psi_{geom}=k\left(\sqrt{L^2+(b+a)^2}-\sqrt{L^2+(b-a)^2}\right)\approx2kab/L.
\end{equation}
The gravitational phase shift is
\begin{equation}
\psi_{gr}\approx\frac{cgm_N^2}{k\hbar^2}\frac{abL}{a+b}
\end{equation}
where $m_N$ is the neutron mass and $c$ is a parameter due to the mirror being tilted with respect to the gravitational field vector. A mirror oriented slightly away from the vertical produces a potential $V_{gr}=cm_Ngz$ where $g$ is the gravitational acceleration and c depends on the angle between the gravitational field vector and the mirror plane. For a mirror tilted $10"$ away from vertical, $c\approx5\times10^{-5}$. The phase shift due to gravity $\psi_{gr}$ may be mitigated by having the neutron mirror positioned as close to vertical as possible. The gravitational phase shift may also be used for calibration by rotating the interferometer about a horizontal axis.

Reflection from the mirror imparts a phase shift $\psi_{refl}$ to the neutron beam of
\begin{equation}
\psi_{refl}\approx\pi-\delta\phi_{refl}
\end{equation}
where
\begin{equation}
\delta\phi_{refl}=2\frac{k_{norm}}{k_b}\approx\pi-2\frac{k}{k_b}\frac{a+b}{L}
\end{equation}
$k_{norm}$ is the component of the neutron wave vector normal to the mirror's surface and $k_b$ is the boundary wave vector of the mirror. The rotation of the Earth contributes a phase shift $\psi_{cor}=(2m_N/\hbar){\bf\Omega A}$ where ${\bf \Omega}$ and {\bf A} are the Earth's angular rotation vector and area enclosed by the interferometer beams vector respectively.

The phase shift $\psi_{cham}$ is the phase shift sourced by the interaction potential from chameleon fields between the neutron beam and the mirror. For a chameleon potential that gives rise to a late time accelerated expansion of the universe: $V=\Lambda^4+\Lambda^{4+n}/\phi^n$, the chameleon-matter interaction potential for a neutron was calculated in \cite{Brax:2011hb} to be
\begin{equation}
V(z)=\beta\frac{m_N}{M_{Pl}\lambda}\left(\frac{2+n}{\sqrt{2}}\right)^{2/(2+n)}\left(\frac{z}{\lambda}\right)^{2/(2+n)}=V_0\left(\frac{z}{\lambda}\right)^{2/(2+n)}
\end{equation}
where $\lambda=\hbar c/\Lambda=82\mu m$ and $\Lambda$ is the dark energy scale. The chameleon matter coupling $\beta$ is contained in $V_0$ as
\begin{equation}
V_0=\beta\;0.9\times10^{-21}\mathrm{eV}\left(\frac{2+n}{\sqrt{2}}\right)^{2/(2+n)}.
\end{equation}
This leads to a chameleon phase shift of
\begin{equation}
\psi_{cham}\approx\frac{\gamma}{\lambda^{\alpha_n-1}\alpha_n}2ab\frac{b^{\alpha_n-1}-a^{\alpha_n-1}}{b^2-a^2}
\end{equation}
where $\alpha_n=(4+n)/(2+n)$ and $\gamma=(m_NV_0L)/(k\hbar^2)$. A neutron interferometer with $a=0.01$cm, $L=1$m, and neutron wavelength $\lambda=100\mathring{A}$, may be sensitive to chameleon-matter couplings below $\beta \sim 10^7$.

Despite the promise of using a Llyod's type interferometer to probe the chameleon-matter coupling, cold neutron interferometry techniques remain yet to be developed and refined. Another type of neutron interferometry experiment described in \cite{Brax:2013cfa} uses a monochromatic beam of slow neutrons split into two coherent beams in an LLL type interferometer. One beam's path through the interferometer is in vacuum, while the other beam's path through the interferometer is in vacuum for part of its path and through a gas with an adjustable density for the remainder of its path. From \cite{Brax:2013cfa}, a neutron beam passing through a region of size $2R$ (in this case a cell filled with gas) with chameleon field $\phi(x)$, will feel an induced potential $m_N\beta\phi(x)/M_{Pl}$ due to the presence of chameleons. The neutron beam passing through the gas experiences a phase shift
\begin{equation}
\psi=\frac{m_N}{k\hbar^2}\int^R_{-R}\frac{m_N\beta}{M_{Pl}}\phi(x)dx.
\end{equation}
For a gas with a low pressure and a large enough chameleon-matter coupling, the chameleons see the gas as a collection of atoms, as opposed to a homogeneous medium. Apart from a small region in the vicinity of the gas atoms and near the walls where the chameleon profile $\phi(x)$ is very steep, the chameleon field has a bubble-like profile between the atoms and walls of the chamber. The slow neutrons possess a large Compton wavelength which allows them to evade screening. As the density of the gas changes, the interatomic distance changes, which in turn changes the chameleon field profile used to calculate the phase shift of the neutron beam. For the neutron interferometry experiment described in \cite{Brax:2013cfa}, the authors find for a beam of slow neutrons ($k=23$nm$^{-1}$) that a gas of Helium at a density of $\rho\approx10$ mbar could be sufficient create a measurable signature in neutron interferometry experiments for chameleon-matter couplings $10^8\lesssim\beta\lesssim 10^{10}$.

\section*{Acknowledgements}
RP would like to thank A. Weltman and W. Wester for useful discussion. RP would like to acknowledge and thank the support of the Claude Leon Foundation.

\end{document}